\documentclass[a4paper,14pt]{article}
\pdfoutput=1 %
\usepackage{jcappub} 
\usepackage[T1]{fontenc} 
\usepackage{color}
\usepackage{subfigure}
\usepackage{graphicx}
\usepackage{bm}

\title{ Circularly Polarized EM Radiation from GW Binary Sources }

\author[a]{Soroush Shakeri}\author[]{}
\author[a]{and  Alireza Allahyari}\author[]{}

\affiliation[a]{School of Astronomy, Institute for Research in Fundamental Sciences (IPM),\\P. O. Box 19395-5531, Tehran, Iran}

\emailAdd{s.shakeri@ipm.ir}
\emailAdd{alireza.al@ipm.ir}

\abstract{ We  consider the  polarization characteristics of the electromagnetic (EM) counterpart of the gravitational wave (GW) created by coalescence of the binary sources. Here, we explore the impact of the photon-graviton interaction on the polarization evolution of X-ray emission of Gamma-Ray Bursts (GRBs). We show that significant circular polarization can be generated due to the gravitational wave from the binary merger.  The circular polarization besides photon energy depends on parameters of GW source such as the chirp mass of the binary, frequency of the GWs and radial distance from the source. Our predicted signal can be used as an indirect probe for GW events and also the nature of photon-graviton interaction. We argue that this polarization signal might be in sensitivity range of upcoming X-ray polarimetry missions.}

\begin{document}
\maketitle
\flushbottom

\section{Introduction}

Detection of the gravitational-wave signal GW170817 by LIGO-Virgo collaboration associated with a short-duration gamma-ray burst (SGRB) observed by the Fermi-satellite, GRB 170817A, has marked the advent of multi-messenger cosmology \cite{GBM:2017lvd, Abbott:2017xzu,Lazzati:2017zsj}. The wealth of science that  joint detections bring is not available from either messenger alone. Thus,  joint detections  can break the degeneracies of the binary parameters and allow to localize the source  of the GW event.
SGRBs are the most promising
EM counterparts associated  with
double neutron star (NS-NS) or neutron-star-black-hole (NS-BH) mergers which are detectable in the sensitivity ranges of both current GW and EM detectors \cite{Metzger:2011bv,Eichler}. In their energy spectrum a high energy prompt emission is followed by X-rays to radio afterglow emissions. These EM counterparts last for seconds to days after the GW event \cite{Metzger:2011bv}. A key question is the origin of the  EM counterparts following the merger. Unambiguous association of GRB signal to GW signal is investigated in Ref \cite{Monitor:2017mdv}. Other counterparts like neutrino emissions associated to the GW event can shed more light on the nature of the merger \cite{ANTARES:2017bia}.

The aim of this article is to introduce a new indicator of the GW events by the  polarization features of its EM  counterpart. We show that the circular polarization of EM emission can serve as a peculiar signature of photon-graviton interaction originating from the binary merger. In this work, we focus on the polarization properties of the very early time  X-ray photons.

Generally, synchrotron radiation as an intrinsic mechanism is applied to explain GRBs polarization, where the level of polarization  depends on  the magnetic field configuration and the geometry of the emission region \cite{Westfold,Mao:2017dlb,Mao:2018rsr}. The  high degree of linear polarization is expected in the presence of an  ordered magnetic field and there are several measurements of linear polarization related to GRB prompt and afterglow emissions \cite{Covino:2016cuw,McConnell:2016lwd}. However, the only  circular polarization measurement belongs to the optical afterglow of GRB 121024A at the level of 0.6 percent \cite{Wiersema:2014bha}. It has been shown that the origin of the observed  circular polarization can not be intrinsic to an optically thin synchrotron process \cite{Nava:2015sba,Matsumiya:2003pw}.
There are few mechanisms to generate circular polarization in the usual scattering processes in the astrophysical medium.
The circular polarization of GRBs can  be generated due to the propagation effects through the conversion of linear polarization of radiation known as Faraday conversion. The generation of the 
circular polarization for GRBs has recently been investigated in Ref. \cite{Batebi:2016efn}, taking into account different types of interactions through the scattering from cosmic particles or being in a background field.  These interactions do not produce huge amounts of circular polarization component.

Additionally, it has been shown that nonlinear QED effects in the strong-field regime lead to  circular polarization in NS \cite{Shakeri:2017knk}. In the case of CMB photons the circular polarization is estimated in Ref. \cite{Montero-Camacho:2018vgs}  and reconsidered by a new approach in \cite{Kamionkowski:2018syl}. The impact of primordial anisotropic background of GWs on the CMB polarization is considered and an unmeasurably low signal is reported in Ref. \cite{Bartolo:2018igk}.  We argue that the considerable values of the circular polarization can be generated by GWs from the binary mergers. On the other hand, we show that the predicted signal might be observable from polarization analysis of the EM counterpart.  More surprisingly, the circular polarization depends on the essential parameters of GW sources such as the chirp mass of the binary, frequency of the GWs, radial distance from the source and moreover the photon energy. 

 Recently, it is shown that  the circularly polarized EM radiation can be generated due to the photon-graviton mixing \cite{Ejlli:2018hke} in which  GWs and EM waves mixing in the presence of external EM fields has been considered. This mixing changes the intensity of both  gravitons and photons. However, we consider photon-graviton forward scattering where the total intensity of photons does not change during interaction with gravitons  

This paper is organized as follows. In section II, we present the formalism of  photon-graviton scattering where  we specialize it to a GW event. In Section III, we briefly present the expressions for GWs from binaries. In section IV, the emission of EM radiation for SGRBs is discussed. We also drive how circular polarization is generated from the binary mergers and provide an order of estimate for the circular polarization. Finally, we conclude in section V.

 \section{Photon Polarization Due to the Photon-Graviton Scattering}\label{do}
  
 Polarization of a light beam is given in terms of Stokes parameters. They quantify the intensity $ I^{\gamma}$, linear polarizations $ Q^{\gamma}, U^{\gamma}$   and circular polarization $ V^{\gamma}$of the beam. 
 The parameters $I^{\gamma}$ and $V^{\gamma}$
   are independent of  the coordinate system and transform as spin-0 (scalar) fields. However,  $Q^{\gamma}$ and $U^{\gamma}$ depend on the orientation of the coordinate system. Meanwhile, $Q^\gamma\pm iU^\gamma$   transform as  spin-2 fields under rotations in the plane perpendicular to the direction of propagation. Note that the total degree of linear polarization $P^{\gamma}=(Q^{2}+U^{2})^{1/2}$ is invariant under rotation of axis. Moreover, Stokes parameters are not Lorentz invariant and transform under Lorentz boosts. It has been shown that $P^{\gamma}/I^{\gamma}$ and $V^\gamma/I^\gamma$ are Lorentz invariant quantities \cite{COCKE}.
Generally the circular polarization is generated by the conversion of linear polarization through Faraday conversion. The generation rate of circular polarization is given by
\begin{eqnarray}
\dot V=2U \frac{d\Delta \phi_{FC}}{dt},
\end{eqnarray}
where $\Delta \phi_{FC}$ is called Faraday conversion phase shift 	\cite{Cooray:2002nm}. The phase shift  $\Delta \phi_{FC}$ is related to the anisotropies in indices of refraction $\Delta n$ which is the property of a birefringent medium \cite{Montero-Camacho:2018vgs}.

Our approach is based on a quantum-mechanical description of polarization  where the time evolution of  Stokes parameters is given by  quantum Boltzmann equation \cite{Kosowsky:1994cy, Alexander:2008fp}. We only consider the leading term (forward-scattering) in the Boltzmann equation. Different types of interactions between a photon beam  and a target beam  can generate  $V$-parameter through the quantum Boltzmann equation. In fact circularly  polarized emission is a result of asymmetry between left- and right-handed circularly polarized photons. 
In principle, circular polarization can be produced by parity violating interactions \cite{Finelli:2008jv,Mohammadi:2013ksa,Mohammadi:2013dea}, intrinsic asymmetric distribution of left- and right- handed components in target beams 	\cite{Vahedi:2018abn,Alexander:2017bxe} and  the presence of an anisotropic  background in the medium \cite{Bavarsad:2009hm,Shakeri:2017knk,Montero-Camacho:2018vgs}

The role of photon-graviton interaction  on the polarization evolution of GRB photons  is the subject of the present paper. It is worth mentioning that gravitons are spin-2 fields under Lorentz transformations. Similar to the case of photons, Stokes parameters $I^{g}$ and $V^{g}$ transform as scalars on the sphere under rotations while  $Q^{g}\pm iU^{g}$ transform as spin-4 fields \cite{Cusin:2018avf}.

The circular polarization due to the photon-graviton scattering is  considered by the authors in the recent work \cite{Bartolo:2018igk}.  The time evolution of Stokes parameters in photon-graviton scattering can be expressed as 
 \begin{eqnarray}
\dot I(\mathbf{k})^{\gamma}=\dot \rho_{11}+\dot \rho_{22}=0,
\label{I}
\label{i12}
\end{eqnarray}

\begin{eqnarray}\label{qdo}
\dot Q(\mathbf{k})^{\gamma} = \dot \rho_{11}-\dot \rho_{22} =
V(\mathbf{k})^{\gamma}\Big(\frac{4 \pi G}{k_{0}}\Big)\int\frac{d^3q}{(2\pi)^3}q^{0} \sin^{2} \theta' \sin 2\phi'~ I(\mathbf{q})^{g},
\label{Q}
\end{eqnarray}

\begin{eqnarray}
\dot  U(\mathbf{k})^{\gamma}= \dot \rho_{12}+\dot \rho_{21} =
-V(\mathbf{k})^{\gamma}\Big(\frac{4 \pi G}{k_{0}}\Big)\int\frac{d^3q}{(2\pi)^3}q^{0} \sin^{2} \theta' \cos 2\phi'~ I(\mathbf{q})^{g},
\label{u}
\end{eqnarray}
\begin{eqnarray}\label{VV2}
\dot V(\mathbf{k})^{\gamma} = i( \dot \rho_{12}-\dot \rho_{21}) =
\Big(\frac{4 \pi G}{k_{0}}\Big) \int\frac{d^3q}{(2\pi)^3}q^{0} \sin^{2} \theta' \Big[\cos 2\phi'~ U(\mathbf{k})^{\gamma}-\sin 2\phi'~ Q(\mathbf{k})^{\gamma} \Big] ~ I(\mathbf{q})^{g},
\end{eqnarray}
where  $\mathbf{q}$ and $\mathbf{k}$ are the momentum vectors of GW and EM radiation  respectively, defined as   
\begin{eqnarray}
\mathbf{q}=q^0(\sin \theta' \cos\phi',\sin \theta' \sin\phi',\cos \theta'), \qquad \mathbf{k}=k^0(0,0,1).
\end{eqnarray}
In these equations superscripts $\gamma$ and $g$ represent photons and gravitons respectively. $\rho_{ij}$ is the density matrix for  photons. 
 We stress that photon-graviton interaction in the presence of an isotropic gravitational wave background can not produce neither linear nor circular polarization. However, photon polarization  can be generated through interaction with an anisotropic distribution of gravitons \cite{Bartolo:2018igk}. Expanding graviton Stokes parameter $I(\mathbf{q})^{g}$ in terms of spherical harmonics as 
\begin{eqnarray}\label{in}
I(\mathbf{q})^{g}=\sum_{lm} a_{lm}Y_{lm}(\mathbf{q}),
\end{eqnarray}
after substituting Eq. (\ref{in}) into Eqs. (\ref{qdo})-(\ref{VV2}), we see that only $Y_{2,\pm 2}$ will give none-zero contributions to the right hand side of these equations. Therefore, the quadrupolar components of the incident intensity distrubution of gravitons leads to generation of circular polarization  from an initially linear polarized radiation. This is consistent with fact that $Q^\gamma\pm i U^\gamma$ transform as  spin-2 fields under rotations.  It is worth mentioning that, in the case of Compton scattering of initially unpolarized photons, linear polarization can be generated only if the radiation intensity has non-zero components $Y_{2,\pm 2}$ \cite{Kosowsky:1994cy}.

The graviton  density matrix for an approximately monochromatic source of a binary merger can be represented by the Dirac delta function as follows
\begin{eqnarray}
\rho_{ij}(\mathbf{q})\propto \delta^3(\mathbf{q}-\mathbf{\bar q}),
\label{mon}
\end{eqnarray}
where $\bar q$ denotes the mean momentum of the gravitational waves produced by a binary merger. In our case gravitons have a preferred direction  acting as  anisotropies in the medium.

As a result, the momentum integral  over  the gravitational wave intensity can be replaced by mean values of the corresponding  quantities of binary merger source.  The mean value of $I(\mathbf{q}) $ can be defined as
\begin{eqnarray}
\int \frac{q^{0} d^3q }{(2\pi)^3}I(\mathbf{q})=\bar I(\bar q).
\label{intensity}
\end{eqnarray}
It is worth mentioning that, the dimension of this quantity is the same as GW energy flux. Using Eqs. (\ref{mon}) and (\ref{intensity}), we can write Eqs (\ref{i12})-(\ref{VV2})  for a monochromatic source as 
 
\begin{eqnarray}\label{II2}
\dot I(\mathbf{k})^{\gamma}=0
\end{eqnarray}
\begin{eqnarray}
\dot Q(\mathbf{k})^{\gamma} = \Omega_{QV}
V(\mathbf{k})^{\gamma},
\end{eqnarray}
\begin{eqnarray}
\dot  U(\mathbf{k})^{\gamma} =
-\Omega_{UV} V(\mathbf{k})^{\gamma}
\end{eqnarray}
\begin{eqnarray}\label{v}
\dot V(\mathbf{k})^{\gamma}  =
\Omega_{UV}~ U(\mathbf{k})^{\gamma}-\Omega_{QV} Q(\mathbf{k})^{\gamma},
\end{eqnarray}
where 
\begin{eqnarray}\label{V12}
\Omega_{QV}=  \Big(\frac{4 \pi G}{k_{0}}\Big)\sin^{2} \theta \sin 2\phi~ \bar I(\mathbf{\bar q})^{g}
\end{eqnarray}
and 
\begin{eqnarray}\label{V22}
\Omega_{UV}= \Big(\frac{4 \pi G}{k_{0}}\Big)\sin^{2} \theta \cos 2\phi~ \bar I(\mathbf{\bar q})^{g}
\end{eqnarray}
According to Eqs. (\ref{II2})-(\ref{V22}), as a result of photon-graviton interaction, the linear polarization of a radiation ($Q^{\gamma}$ and/or $U^{\gamma}\neq 0$) can be converted into the circular polarization ($V^{\gamma}\neq 0$) in the presence of nonzero GW energy flux ($\bar I^{g}\neq 0$).

To investigate the transformation properties of Eq. (\ref{v}), we use the approach presented in \cite{Kamionkowski:2018syl}. Thus, we introduce the matrices
\begin{equation}
     \mathcal{P}_{ab}(\mathbf{k}) = \frac{1}{\sqrt{2}}\left( \begin{matrix}
     Q^{\gamma}(\mathbf{k}) & U^{\gamma}(\mathbf{k}) \\ U^{\gamma}(\mathbf{k}) &
     -Q^{\gamma}(\mathbf{k}) \end{matrix} \right),
\end{equation}
and
\begin{equation}
     \Phi_{ab}(\theta, \phi) = \frac{1}{\sqrt{2}}\left(\begin{matrix}
     \Omega_{UV}(\theta, \phi) &
     \Omega_{QV}(\theta, \phi) \\ \Omega_{QV}(\theta, \phi) &
     -\Omega_{UV}(\theta, \phi) \end{matrix} \right).
\label{eqn:phitensor}     
\end{equation}
Then we can re-write Eq. (\ref{v}) as 
\begin{equation}\label{vdot}
     \dot V^\gamma(\mathbf{k})= \epsilon_{ac} \mathcal{P}^{ab}(\mathbf{k}) \Phi_{b}^{\ c}(\theta, \phi),
\end{equation}
where $\epsilon_{ac}$ is the antisymmetric tensor on the sphere. Regarding $U$ and $Q$ rotation transformations \cite{Kosowsky:1994cy}, and Eqs. (\ref{V12}), (\ref{V22}),  one can show that under rotations  ${ \bm{\mathcal{P}}}$ and ${\bf \Phi}$ matrices transform as
\begin{align}\label{pp1}
 \bm{\mathcal{P'}}= \left(\begin{matrix}
\cos{2 \varphi} & \sin{2 \varphi}\\-\sin{2 \varphi} & \cos{2 \varphi}\end{matrix}\right){\bm{\mathcal{P}}},
\end{align}
\begin{align}\label{pr}
{\bf \Phi'}= \left(\begin{matrix}
\cos{2  \varphi} & -\sin{2  \varphi}\\\sin{2  \varphi} & \cos{2  \varphi}\end{matrix}\right){\bf \Phi}.
\end{align}
By using Eqs. (\ref{vdot}), (\ref{pp1}) and (\ref{pr}), it is clear that $V$ transforms as a spin-0 field under rotations. In the following  sections we consider GW from binary mergers and EM radiation from SGRBs where we will discuss the GWs and EM radiation interaction. We will relate $\bar I(\mathbf{\bar q})^{g}$ to GW energy density of the binary merger. This yields the generation rate of circular polarization due to GWs from the binary merger.

 \section{GW Radiation from Inspiral of Compact Binaries}
\begin{figure}
    \centering
    {
        \includegraphics[width=3.8in]{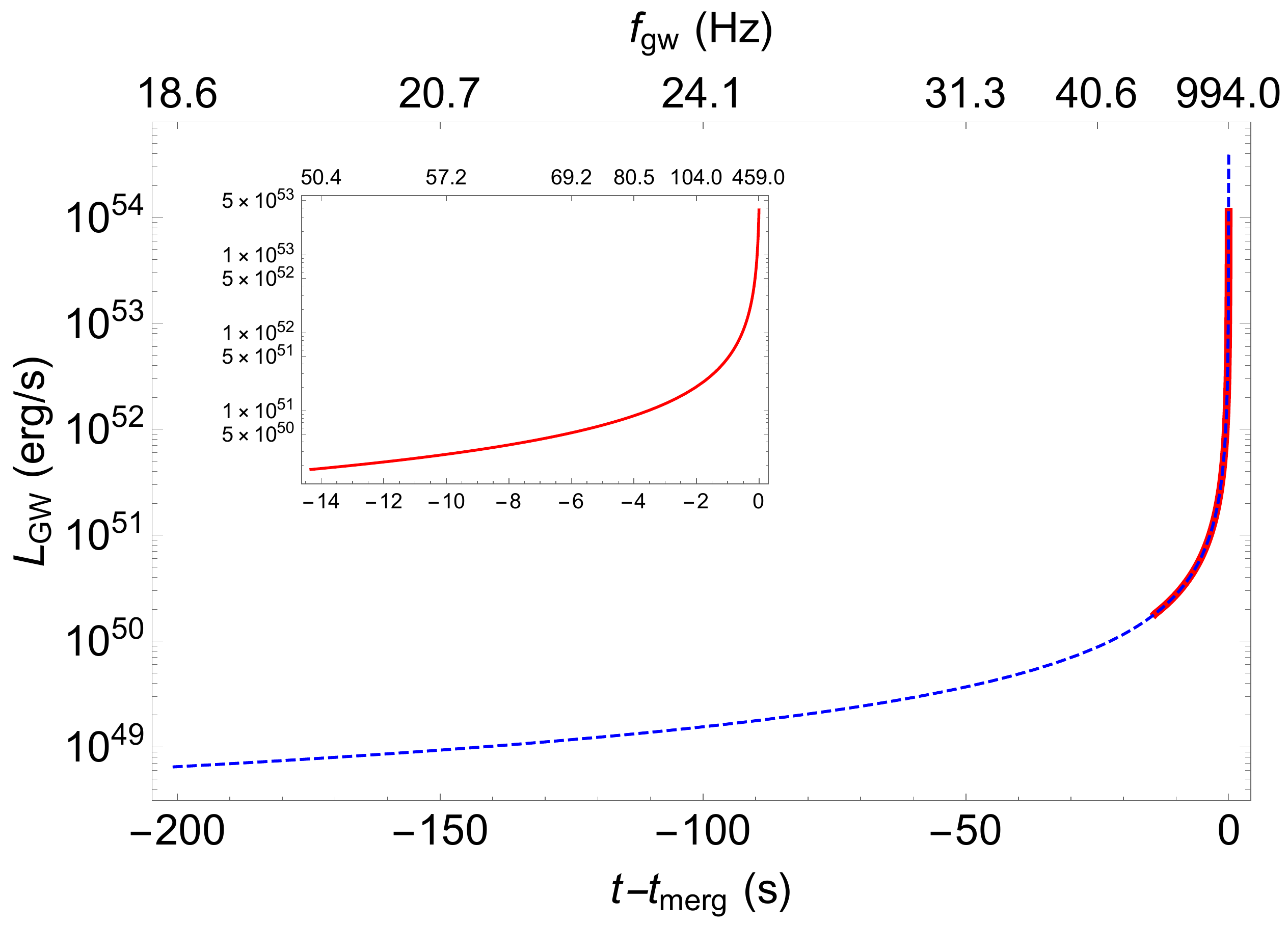}
        }
    \caption{The gravitational wave luminosity due to the binary merger as a function of time from the merger. We used the quadrupole approximation with the chirp mass  $M_{c}=1.188 M_{\bigodot}$ and  the total binary mass $M=2.82 M_{\bigodot}$  in the frequency range $f_{gw}\simeq18.6-1000 Hz$ [dashed (blue)]. The frequency range $f_{gw}\simeq50-460 Hz$ [solid (red)] is in agreement  with GW170817 observed by LIGO-Virgo collaboration, where the insert shows this range in a larger scale.}
    \label{Fluxm}
\end{figure}

 Here we consider gravitational radiation due to a binary system made of two compact stars, such as neutron stars or black holes \cite{Abbott:2016bqf,Maggiore:2018tk}. In this case the majority of gravitational radiation will be emitted  at the twice of the orbital frequency ($\omega_{gw}=2\omega_{s}$), where $\omega_{s}=GM/R^3$. In general the emission of GWs causes the orbital radius R to decrease in time and therefore $\omega_{s}$ increases. Hence, the time-dependence of  $\omega_{gw}$ is as follows
 
 \begin{eqnarray}\label{ws}
w_{gw}=2\pi f_{gw}= 2 (\frac{5}{256}\frac{1}{\tau})^{3/8}(\frac{GM_{c}}{c^{3}})^{-5/8},
\end{eqnarray}
where $\tau =t_{merg}-t$ (the time to merger) and the chirp mass 
 $M_{c}=\mu^{3/5} M^{2/5}$ is a function of the reduced mass $\mu= m_{1}m_{2}/(m_{1}+m_{2})$ and the total mass $M=m_{1}+m_{2}$ of the binary system.
  The power at solid angle $d \Omega$ is $\frac{d P}{d \Omega}$. Thus, the GW  energy density   $\rho_{gw}=\frac{1}{cr^2 }\frac{d P}{d \Omega }$ in the quadrupole approximation is given by
 
\begin{align}\label{gw1}
\rho_{gw}=\frac{2}{\pi r^2}\frac{c^4}{G}(\frac{G M_c \omega_{gw}}{2 c^3})^{\frac{10}{3}}g(\theta),
\end{align}
where $g(\theta)=\cos^2\theta+\left( \dfrac{1+\cos^2\theta}{2}\right)^2 $  \cite{Maggiore:2018tk}. The gravitational wave luminosity or total radiated power can  be expressed in terms of the mass and period as
 \begin{eqnarray}
L_{gw}&=& \int c \rho_{gw}r^{2}d\Omega=
\frac{32}{5}\frac{c^5}{G}\Big(\frac{GM_{c}\omega_{gw}}{2c^3}\Big)^{10/3}
\\ \nonumber 
 &=& 2.48\times10^{51} \Big( \frac{M_{c}}{M_{\bigodot}}\Big)^{10/3} \Big( \frac{f_{gw}}{100Hz}\Big)^{10/3} erg \ s^{-1}.
\end{eqnarray}
For GW170817 with the binary chirp mass $M_{c}=1.188_{-0.002}^{+0.004} M_{\bigodot}$, the total binary mass $M=2.82_{-0.09}^{+0.47} M_{\bigodot}$  corresponds to the case of equal mass components $m_{1}=m_{2}=1.365 M_{\bigodot}$ \cite{Monitor:2017mdv}, the GW luminosity in different frequency ranges  is displayed in Fig \ref{Fluxm}. This plot shows that the GW luminosity  increases substantially  up to $10^{54}erg/s$ before the merger. Note that the ground-based interferometers, such as LIGO and Virgo are designed to detect gravitational waves in the frequency range from 10 Hz to 10 kHz \cite{Martynov:2016fzi}.
The GW energy flux at the distance r from the source is 
 \begin{eqnarray}
\mathcal{F}_{gw}\approx \frac{L_{gw}}{4\pi r^2}.
\end{eqnarray}
We usually observe this flux on the ground and it can be traced back to the source. 

\section{EM Radiation from SGRBs}
In order to consider GW-EM radiation interaction, we need to specify the EM emission source and also justify how the EM radiation could interact with the GW during the merger. The  GRB emissions in different energy bands  come from different locations with respect to the  burst   progenitor. Here we focus on the SGRBs with $T_{90}<2s$ with the binary progenitor. Independent of the burst progenitor, it is assumed that an outflow including $e^{+}e^{-}$ pairs coexist with photons in thermal equilibrium.  This initially thick plasma  keeps accelerating until reaching transparency \cite{Meszaros:1999gb,Lyutikov:2003ih,Nakar:2007yr,Piran:2004ba,Ruffini:2009hg}.
The outflowing materials originate from  radius $R_{0}=6GM_{BH}/c^{2}$ which is  the innermost stable circular orbit of a Shwarzchild BH with a mass equal to $M_{BH}$, this value is  equal to $2.5\times10^{6}cm$ for $M_{BH}=2.8 M_{\bigodot}$ \cite{Meszaros:1999gb}.

Generally, the GRBs energy spectrum comprising  both the thermal and non-thermal parts of the spectra vary with time. 
In the regime of high optical depth ($\tau\gg1$) which is related to the early time emission, all emerging radiation must be thermal. Since both EM flux and luminosity are observable quantities, the emitter radius $r_{em}$ due to
the blackbody emission is 
 \begin{eqnarray}\label{radi2}
r_{em}\approx \sqrt{\frac{L_{EM}}{4\pi \mathcal{F}_{EM}}}=\sqrt{\frac{L_{EM}}{4\pi \sigma T^{4}}}=2.7\times 10^{9}  \Big(\frac{L_{EM}} {10^{47}erg/s} \Big)^{1/2}\Big( \frac{10 keV} {T}  \Big)^{2} cm.
\end{eqnarray}
This approximate formula  can be modified by applying relativistic corrections \cite{Kumar:2014upa,Ruffini:2017xsr}. The isotropic  energy of SGRBs usually ranges between $10^{49}-10^{51}$ erg \cite{Nakar:2007yr}, in GRB170817  the isotropic energy of soft thermal blackbody (BB) component, $E_{iso,BB}=1.3\times10^{46}$ erg, and the isotropic peak luminosity, $L_{iso,p}=1.6\times10^{47}$ erg/s \cite{Monitor:2017mdv}. 

Meanwhile, the EM emission during the premerger stage of  binary NSs  may be produced by the preburst activity in SGRBs \cite{Palenzuela:2013hu,Berger:2013jza}, this leads to photons emitted  from a radius  smaller than (\ref{radi2}). Such EM precursors would be generated due to the magnetospheric interaction of a binary NS \cite{Troja:2010zm,Hansen:2000am} or resonant shattering of the NS crusts \cite{Tsang:2011ad} prior to merger. It has been considered the possibility of the energy extraction  of about $10^{45}-10^{46}$ erg/s  which is deposited into magnetospheric pair plasma at $r\approx10^{7} cm$ \cite{Hansen:2000am}.
Precursors  1-10 s prior to the main burst were detected with high significance for three cases out of the 49 SGRBs considered in Ref. \cite{Troja:2010zm}.  Therefore it is useful to study the EM emission  before the merger,  to correlate it with the emitted GWs, and to examine whether the photon-graviton  interaction can yield observable signatures on EM polarization signals. Such EM precursors might provide an ideal probe for photon-graviton interaction in our scenario.

In fact it is unclear whether a jet should be always produced
after the binary merger. In our scenario the GWs  might also pass through the optically thick plasma and the trapped photons have enough time  to interact with gravitons to produce circular polarization. Additionally the matter-rich environment surrounding the neutron star leads to  multiple scatterings of photons. This even gives more time to photons to interact with the GWs. Consequently, the GRB emission might be generated simultaneously with the GW events and reach us with a delayed arrival time. At the photospheric radius $r_{ph}$, where the optical depth $\tau=1$, the photons decouple from the plasma  and start streaming freely.  The delay between the GW signal and GRB signal can also be interpreted as the time it takes the photosphere to radiate.



\subsection{Photon circular polarization}
\begin{figure}
    \centering
        \subfigure
    {
        \includegraphics[width=2.9in]{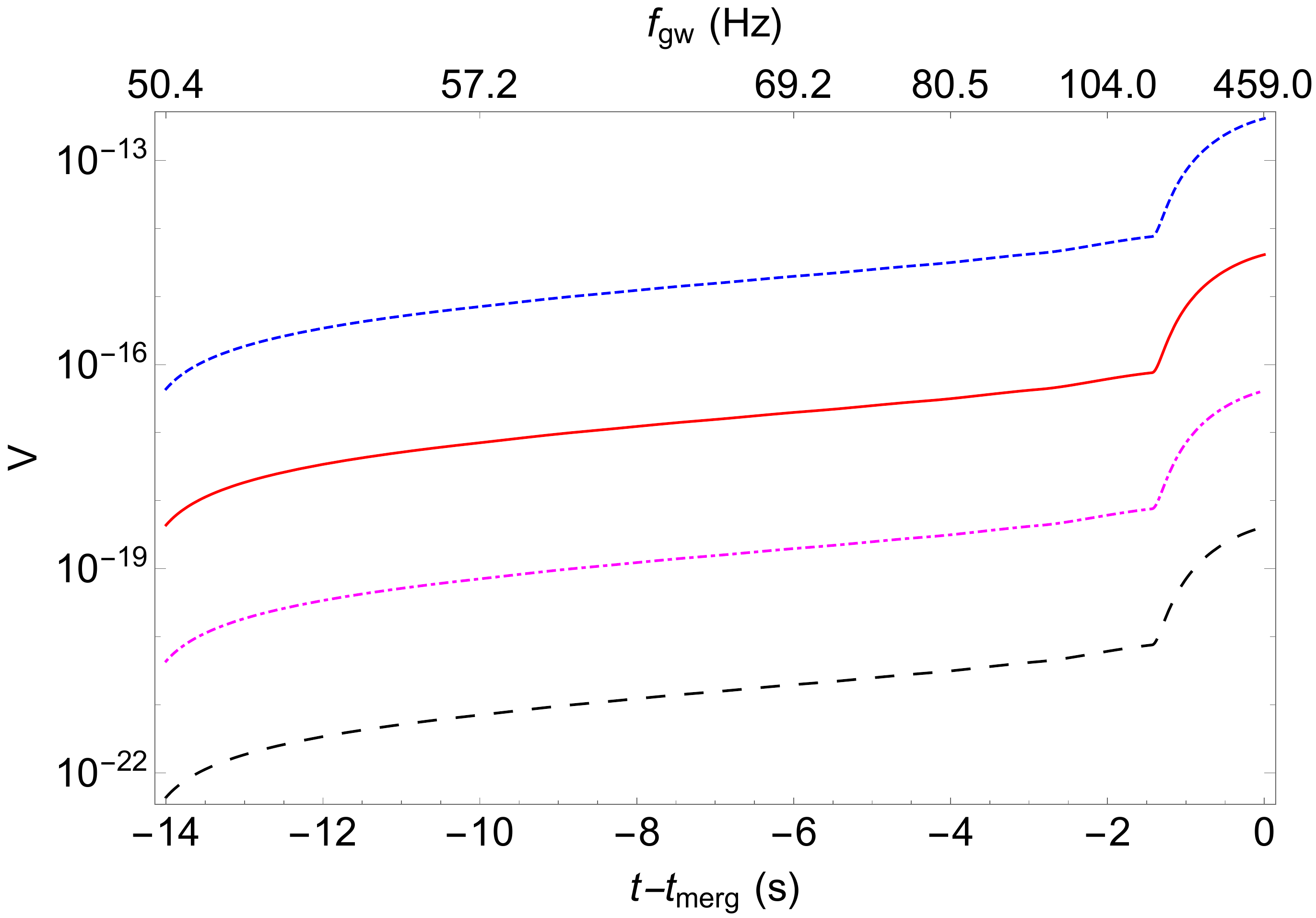}        
     }  
       {
        \includegraphics[width=2.85in]{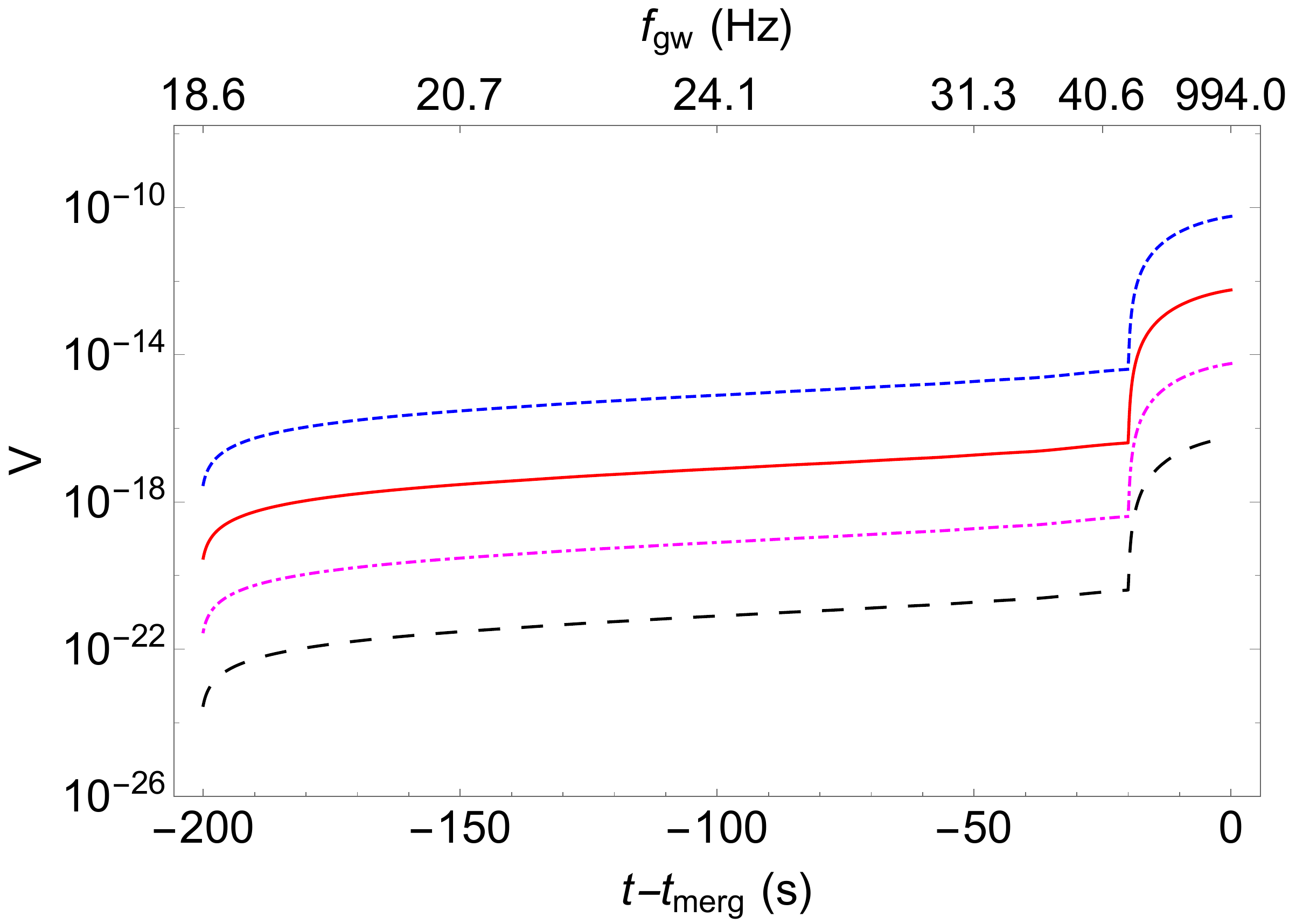}        
     }  
    \caption{The time evolution of  circular polarization parameter due to the interaction with  GWs at the different radial distances from the binaries, $r=10^6 cm$ [dashed(blue)], $10^7 cm$ [solid(red)], $10^8 cm$ [dot-dashed(magenta)] and $10^9 cm$ [long dash(black)]. We plotted dimensionless Stokes parameter V normalized by the photon intensity as a function of time form the merger.
    \textit{Left}:
     The circular polarization  in the last $14 s$ and in the frequency range $f_{gw}\simeq50-460 Hz$.
   \textit{Right}: The circular polarization  in the last $200 s $ from the merger and in the frequency range $f_{gw}\simeq18.6-1000 Hz$. The time evolution of GW luminosity during these time intervals is given in Fig. \ref{Fluxm}.}
    \label{f01}
    \end{figure}
The Compton scattering and its inverse process  are the most dominant scattering processes in both the GRB shockwaves and in the interstellar medium (ISM).  Although these processes can be the original  source of the linear polarization, they cannot generate circularly polarized emissions.  In Sec. \ref{do}, we showed that the photon-graviton scattering behaves as a source term for circular polarization presented in Eq. (\ref{v}).
Let us first give an order of estimate for circular polarization due to the GW-EM radiation interaction
 \begin{align}\label{vd}
\dot{V}^{\gamma}\approx \frac{4\pi G}{k_{0}} \mathcal{F}_{gw}U^{\gamma}=5.7\times10^{-19}\Big( \frac{L_{gw}}{10^{51}erg/s}\Big)\Big( \frac{10^{9} cm}{r}\Big)^{2}\Big( \frac{10 keV}{k_{0}}\Big) U^{\gamma} \; s^{-1},
\end{align}
This formula shows the rate of circular polarization generation for a binary merger. In Eq. (\ref{vd}), r is the radial distance from the binary source,  where the GW-EM interaction is expected to occur. For the maximum luminosity value of GW170817, $L_{gw}=10^{54}erg/s$ and the typical radius  $r=10^{6} cm$, $\dot{V}^{\gamma}/U^{\gamma}$ increase up to $10^{-10} s^{-1}$ for $10 keV$ photons. The typical GW peak luminosity emitted by compact binary mergers is $3.63 \times (10^{56}-10^{55})  erg/s$ \cite{Zappa:2017xba}, so  $\dot{V}^{\gamma}/U^{\gamma}$ can be increased up to $10^{-7} s^{-1}$ taking into account the maximum value of GW peak luminosity. Thanks to recent technological advances in X-ray polarimetry \cite{Marx:2013xwa,Bellazzini:2010rw}, the X-ray polarimeters allow detection of rotation of the polarization plane ($1/2\tan^{-1}(U^{\gamma}/Q^{\gamma})$) down to 1 arcsec ($4.8\times 10^{-6}$ rad) and ellipticities ($1/2\tan^{-1}(V^{\gamma}/P^{\gamma})$) of about $1.51\times 10^{-5}$ rad \cite{Shakeri:2017iph}. The limited duration and isotropic distribution of GRBs, requires a polarimeter with a large field of view and high sensitivity. The acquired statistics for the bright GRBs allows for detailed polarization measurements.

To relate $\bar I^{g}$ to the binary parameters, we write it in terms of the GW energy density $\bar I^g(\bar q)=4  \rho_{gw}$  \cite{Bartolo:2018igk}. Therefore we can rewrite  Eqs. (\ref{V12})-(\ref{V22}) as
\begin{align}\label{qv1}
\Omega_{QV}=-\dfrac{ 32 c^{5}}{ k_0 r^2} (\frac{G M_c \omega_{gw}}{2 c^3})^{\frac{10}{3}}\sin^2\theta  \sin 2\phi \ g(\theta),
\end{align}
and 
\begin{align}\label{qv2}
\Omega_{UV}=\dfrac{ 32 c^{5}}{ k_0 r^2} (\frac{G M_c \omega_{gw}}{2 c^3})^{\frac{10}{3}}\sin^2\theta  \cos 2\phi \ g(\theta).
\end{align}
Regarding the set of Eqs. 
 (\ref{II2})-(\ref{v}), in the case  of  time-independent orbital frequency $\omega_{s}$ of the binary,   Eq.  (\ref{v}) becomes a simple harmonic equation $\ddot{V}+\Omega^2_{v} V=0$. The frequency of  oscillations is given by 
\begin{eqnarray}\label{4r}
\Omega_{v} = 1.97\times 10^{-21}\Big( \frac{M_{c}}{M_{\bigodot}}\Big)^{10/3} \Big( \frac{f_{gw}}{100Hz}\Big)^{10/3} \Big( \frac{10^{9} cm}{r}\Big)^{2}\Big( \frac{10 keV}{k_{0}}\Big) \sin^2\theta \ g(\theta) \ s^{-1}.
\end{eqnarray}    
Generally $\omega_{gw}$ is time dependent as given by Eq. (\ref{ws}). Regarding Eqs. (\ref{qv1}) and (\ref{qv2}), we solve Eqs.  (\ref{II2})-(\ref{v}) numerically taking into account $\omega_{gw}$ as a time-dependent function and  by fixing the parameters as, $M_c=1.188 \ M_{\bigodot}$, $k_0=10 \ keV$, $\theta=\pi/4$ and $\phi=\pi/8$.
The time evolution of circular polarization for different values of $r$ parameter is displayed  in Fig. \ref{f01}, as a function of time from the merger. These diagrams are plotted as a function of the merger duration since  GW luminosity is an increasing function of time to the merger. Since GRB photons come from different radial distances from the binary source, Fig. (2) shows that at the specific radius how much circular polarization can be produced due to photon-graviton interaction. Meanwhile, we supposed a totally linear polarized radiation $U_{0}=1$ and $Q_{0}=0$ without any initial circular
polarization $V_{0}= 0$, which are all dimensionless
quantities.  The Stokes parameters are normalized by the photon intensity $I^{\gamma}$, this yields  the dimensionless quantities. The left panel in Fig. \ref{f01},  shows the time evolution of the circular polarization for a merger that lasts $14 s$ in  the frequency range $f_{gw}\simeq50-460 Hz$. The right panel shows the time evolution of the circular polarization for a merger that lasts $200 s$ in the frequency range $f_{gw}\simeq18.6-1000 Hz$.  The value of circular polarization can rise up to $10^{-10}$ for X-ray photons at $r\approx 10^6 cm$. The value of the circular polarization depends on both  the duration of interaction and the luminosity of the GW at the time of interaction. Our results show that even for short interaction time,  the generation of large values of circular polarization is possible.

Our obtained values are considerably large with respect to the values generated by other mechanisms discussed in the introduction; e.g. the mechanisms proposed in  Ref. \cite{Batebi:2016efn}.
According to the angular dependence of Eqs. (\ref{qv1}) and (\ref{qv2}),  the circular polarization gets its maximum rate at $\theta \approx \pi/4$ and $\theta \approx 3\pi/4$ and the zero rate is obtained at $\theta =0$ and $\theta =\pi$. Propagation of photons and gravitons at exactly the same direction towards the earth ($\theta=0$) can not produce any EM polarization signals due to photon-graviton interaction. Therefore, in our scenario the polarization changes due to  photon-graviton interaction occur only during the merger. Note that in order to have considerable values of circular component, it is not essential to have long duration of interaction time between gravitons and photons. As it is shown in Fig. \ref{f01} (left panel), most of the circular polarization is produced in less than 2$s$ to merger time. Although we fix the collision angle between photons and gravitons during the interaction, the photon multiple scatterings change the photon direction many times in the outflowing materials. Therefore, the observed photons along the $\mathbf{q}$ direction are not necessarily at this direction during the photon-graviton interaction.  One can assume that gravitons interact  with photons during merger time and photons at the interaction point have different directions with respect to the gravitons.Generalization of our computation to a more realistic configuration taking into account  variations in  photon direction due to other scattering processes will be the subject of a future paper.

Around the frequency 100 Hz, the LIGO and Virgo are in their most sensitive band which is in the level of $10^{-23} (Hz)^{-1/2}$. However, shot noise limits their sensitivity at frequencies  above 100Hz \cite{Martynov:2016fzi}. One of the advantages of polarization analysis of the EM counterpart is giving  rich informations even outside the sensitivity range of the current GW detectors. As is shown in Fig. \ref{f01}, photons obtain  most of their values of circular polarization when interacting with higher GWs  luminosity  near the merger. This makes circular polarization  an indirect probe for GWs even when  GW detection is not possible by our  detectors.

\section{Conclusion and Remarks}
In this paper, we have investigated the circular polarization induced on EM radiation accompanying gravitational waves from binaries. We use the  quantum Boltzmann equation for the evolution of the polarization, taking into account photon-graviton interaction. We proposed a scenario for the interaction of GRB photons with gravitational waves due to the binaries. Using the relevant parameters of recent observations for the GRB-GW event, we solved the set of equation describing the evolution of Stokes parameters. Here, we considered X-ray  photons of SGRBs. The photon polarization has a characteristic dependence on binary parameter as the GW source. We conclude that the substantial amount of circular polarization can be generated due to the GW-EM interaction in binaries.  Hence, the circular polarization associated to a GW merger might be observed in the prospective X-ray polarimetry experiments. Although our scenario is based on some simple assumptions, it captures novel aspects of the polarization signals due to photon-graviton interaction in the EM counterpart of GWs.

Precise measurement of GRB polarization is one of the main goals for future GRB observations which can yield valuable information about the radiation mechanisms and the photon interactions, especially photon-graviton interaction.  Several polarimeters are expected to be launched within several years \cite{Bellazzini:2010rw,Costa:2001mc,Bellazzin2,Soffitta2,Soffitta3}, such as the X-ray Imaging Polarimetry Explorer (XIPE), the Imaging X-ray Polarimetry Explorer (IXPE) and the Polarimeter for Relativistic Astrophysical X-ray Sources (PRAXyS). These instruments might  be able to detect the polarization signals from photon-graviton interaction  accompanying binary mergers and can shed light on the nature of this interaction.

\section*{Acknowledgements}  
S. Shakeri would like to thank R. Ruffini for supporting his visit at ICRANet Pescara and Rome, where the last part of this work was done. He is also grateful to Y. Wang for his useful comments which substantially helped to improve this article. S. Shakeri and A. Allahyari  are grateful to H. Firouzjahi and S. Matarrese for reading the preliminary manuscript and insightful comments and also to M. Zarei and R. Mohammadi for useful comments during revision process.

\end{document}